\documentclass[11pt]{article}
\usepackage{latexsym, epsfig, verbatim, amsmath, amssymb, lscape, graphics, threeparttable, color}
\usepackage{mathtools, multirow, rotating}
\usepackage[margin = 1in]{geometry}
\usepackage{setspace}
\usepackage{authblk}
\usepackage{hyperref}

\usepackage{latexsym, epsfig, verbatim, amsmath, amssymb, lscape, graphics, threeparttable, color}
\usepackage{mathtools, multirow, rotating}
\usepackage{enumitem}
\usepackage[labelfont=bf]{caption}
\usepackage[utf8]{inputenc}
\usepackage[english]{babel}

\allowdisplaybreaks
\newtheorem{remark}{Remark}

\newtheorem{theorem}{Theorem}

\newcommand{\bgamma}{{\boldsymbol \gamma}}

\newcommand{\wgamma}{{\widehat\gamma}}
\newcommand{\wbgamma}{{\widehat\bgamma}}
\newcommand{\wlambda}{{\widehat\lambda}}

\newcommand{\wbeta}{{\widehat\beta}}
\newcommand{\wmu}{{\widehat\mu}}
\newcommand{\wOmega}{{\widehat\Omega}}
\newcommand{\wphi}{{\widehat\phi}}

\newcommand{\tlambda}{{\widetilde \lambda}}

\newcommand{\calR}{{\cal R}}

\newtheorem{assumpsnap}{Assumption}

\newtheorem{assumpkan}{Assumption}

\newtheorem{assumpC}{Assumption}

\newtheorem{assumpD}{Assumption}

\newtheorem{assumpE}{Assumption}

\usepackage{natbib}
\usepackage[superscript]{cite}

\begin{document}

\title{Statistical Considerations for Cross-Sectional HIV Incidence Estimation Based on Recency Test}
\author[1,2]{Fei Gao*}
\author[3]{Marlena Bannick}

\affil[1]{Vaccine and Infectious Disease Division, Fred Hutchinson Cancer Research Center, Seattle, WA}

\affil[2]{Public Health Sciences Division, Fred Hutchinson Cancer Research Center, Seattle, WA}

\affil[3]{Department of Biostatistics, University of Washington, Seattle, WA}
\date{}
\bigskip
\maketitle
\begin{abstract}
Longitudinal cohorts to determine the incidence of HIV infection are logistically challenging, so researchers have sought alternative strategies.
Recency test methods use biomarker profiles of HIV-infected subjects in a cross-sectional sample to infer whether they are ``recently'' infected and to estimate incidence in the population.
Two main estimators have been used in practice: one that assumes a recency test is perfectly specific, and another that allows for false-recent results.
To date, these commonly used estimators have not been rigorously studied with respect to their assumptions and statistical properties.
In this paper, we present a theoretical framework with which to understand these estimators and interrogate their assumptions, and perform a simulation study to assess the performance of these estimators under realistic HIV epidemiological dynamics.
We conclude with recommendations for the use of these estimators in practice and a discussion of future methodological developments to improve HIV incidence estimation via recency test.\end{abstract}

\raggedbottom

\footnotetext{\textbf{Abbreviations:} FRR, false-recency rate; MDRI, mean duration of recent infection}

\section{Introduction}

Determination of the incidence rate of HIV is critical for HIV surveillance and for evaluating the effectiveness of HIV prevention efforts.
The current gold standard is through longitudinal follow-up and repeated testing of a cohort of participants drawn from the population of interest, such that the incidence can be estimated by the ratio of number of new cases and total follow-up time.
This approach is theoretically simple, but it may present issues for HIV surveillance \citep{brookmeyer2010measuring}.
For example, high follow-up rates in large representative samples may be difficult to obtain, the cost of such studies is usually high, and there may be differences in HIV risk behaviors among persons who do and do not participate in cohort studies.
In addition, there may be retention bias during follow-up or alteration of HIV risk by repeated HIV counseling and testing (Hawthorne effect) \citep{sherr2007voluntary}.

An important alternative approach that avoids longitudinal follow-up and repeated testing is cross-sectional incidence estimation.
This approach utilizes a biomarker-based algorithm to determine which infections in a cross-sectional sample drawn from the population of interest were acquired ``recently''.
It was first proposed by \citet{brookmeyer1995estimation}, where subjects with negative HIV-antibody test and positive HIV-1 p24 antigen test were classified as recently infected.
This recency test is infeasible in practice since the short p24 antigen-positive pre-seroconversion period requires testing a large number of individuals to estimate incidence with precision.
Later, a number of serological assays that measure the antibody response to HIV infection were proposed, for example the detuned assay \citep{janssen1998new}, the BED capture EIA  \citep{parekh2002quantitative}, and the avidity assay \citep{suligoi2002precision, duong2015recalibration}.
Genetic diversity of HIV has also been used as a biomarker to indicate HIV recency \citep{kouyos2011ambiguous,yang2012new, cousins2011use}.
To improve the performance of the recency algorithm, others have proposed multi-assay algorithms that make use of multiple assays and biomarkers to indicate recency \citep{brookmeyer2013estimation, laeyendecker2013HIV, konikoff2013performance,laeyendecker2018identification}.

A number of statistical approaches have been proposed to determine HIV incidence based on a cross-sectional sample.
They make use of recency test results in a cross-sectional sample as well as the classification characteristics of the recency test.
Based on the understanding that incidence can be viewed as the expected number of new infections per uninfected person per unit time, \citet{kaplan1999snapshot} proposed the ``snapshot estimator''
\begin{equation}
    \wlambda = \frac{N_{rec}}{\wmu N_{neg}},\label{equ:snapshot}
\end{equation}
where $N_{rec}$ and $N_{neg}$ are the numbers of test-recent HIV-positive subjects and HIV-negative subjects, respectively, from the cross-sectional sample, and $\wmu$ is an estimate of the ``mean window period'' of the recency test (the average duration of infection among the subjects classified as recently infected; we will discuss a formal definition of this parameter in Section \ref{sec:meth}).
The snapshot estimator was suggested to be unbiased when the incidence is constant over time, and it has been adopted in a number of applications in HIV incidence estimation \citep{eshleman2013use,rehle2015comparison,solomon2016community}.

The snapshot estimator implicitly assumes that the mean window period is finite, such that a long-infected subject would have a zero probability of being classified as recent.
However, for many recency tests, a small proportion of long-infected persons may be falsely classified as ``recent''.
A number of methods have been proposed to address such false-recency \citep{mcdougal2006comparison,hargrove2008improved, kassanjee2012new}.
One widely adopted approach is the ``adjusted estimator'' from \citet{kassanjee2012new}, where an infection duration cutoff $T^*$ is defined to delineate between ``recent'' and ``long'' infected subjects.
Based on this cutoff, the adjusted estimator uses two characteristics of a recency test that are closely related to the sensitivity and specificity of a classification procedure: mean duration of infection (MDRI) $\Omega_{T^*}$ and false-recent rate (FRR) $\beta_{T^*}$.
MDRI is similar to the mean window period in that it captures the average duration of infection among those who are ``truly recent'' and classified as recently infected, and FRR is the probability of mis-classified for a randomly selected long-infected subjects (these parameters will be further discussed in Section \ref{sec:meth}).
For a recency test, the adjusted estimator is given by
\begin{equation}
    \tlambda = \frac{N_{rec} - N_{pos}\wbeta_{T^*}}{N_{neg}(\wOmega_{T^*} - \wbeta_{T^*}T^*)},\label{equ:kassanjee}
\end{equation}
where $N_{pos}$ is the number of HIV-positive subjects in the cross-sectional sample, and $\wOmega_{T^*}$ and $\wbeta_{T^*}$ are estimates of MDRI and FRR for the recency test, respectively.
Since it accounts for recency tests that produce false-recent results, the adjusted estimator is thought to be more flexible and theoretically more robust than the snapshot estimator.
It has also been widely adopted in applications of HIV incidence estimation \citep{maman2016closer,moyo2018cross}.

Even though the cross-sectional incidence estimators have been widely utilized in practice, the statistical properties, especially those of the adjusted estimator, have not been well studied or understood.
Specifically, the key parameters (MDRI, FRR) may not be well characterized and the assumptions under which the estimators serve as unbiased estimators for the incidence in a target population have not been rigorously studied.
In this paper, we formulate a theoretical framework for assessing HIV recency, formally establish the assumptions for cross-sectional incidence estimation based on the snapshot and adjusted estimators, and evaluate the bias of the estimators when the assumptions fail to hold.
We evaluate the numerical performance of the estimators under various simulated settings with different HIV epidemic trajectories and recency tests with different properties and provide some recommendations in using the estimators in practice.

\section{Theoretical Model}
\subsection{Notation}
Let $T$ be the (calendar) HIV infection time of a subject and let $A(t)$ be an indicator of eligibility at time $t$, i.e., whether this subject would be eligible to be included in a survey for HIV incidence (and prevalence) of the target population at time $t$.
This eligibility indicator $A(t)$ can be based on a collection of (possibly time-dependent) individual covariates and a population of interest.
For example, in a cross-sectional population survey, a minimum requirement for eligibility is being alive at the time the survey is conducted.
Another example is $A(t) = I(\text{MSM, Age 18-50 at time }t)$, which uses two characteristics -- an indicator of being a member of the men who have sex with men population (MSM), and an indicator of being aged 18-50 -- to define the eligible population at time $t$.

At any calendar time $t$, the prevalence in this target population is given by $p(t) =\Pr(T\le t | A(t)=1)$, and the incidence in this population is given by 
\begin{equation}
    \lambda(t) = \lim_{dt\to0}\frac{1}{dt} \Pr(t\le T<t+dt | T\ge t,A(t)=1),\label{equ:inc}
\end{equation} 
i.e., it is the rate of instantaneous HIV infection for an eligible HIV-negative subject at time $t$.
With slight abuse of notation, we write $\lambda(t) = \Pr(T=t|T\ge t,A(t)=1)$.
Note that both $p(t)$ and $\lambda(t)$ concern the distribution of infection time $T$ in a restricted population defined by $A(t)$, such that they are conditional quantities given $A(t)=1$.
The main goal of cross-sectional incidence estimation is to estimate $\lambda(t)$ based on a finite-sized cross-sectional sample collected at time $t$ satisfying $A(t) = 1$.

\subsection{Cross-Sectional Sample}\label{sec:meth}
Suppose that we collect a random sample from the eligible population at time $t$ (all subjects in the sample satisfy $A(t)=1$).
For each subject in that sample, we first assess their HIV status, and if it is positive, we apply a subsequent HIV recency test.
We assume that HIV status can be determined without any mis-classification (e.g., using an RNA-based diagnostic), while HIV recency may not be, with details described below.
Recall that we use $N_{neg}$, $N_{pos}$, and $N_{rec}$ to denote the numbers of subjects who are HIV-negative, HIV-positive, and HIV test-recent, respectively.
The probabilities associated with those subjects in the cross-sectional sample are:
\begin{itemize}
\item The probability of HIV-negative: $\Pr(T>t | A(t)=1) = 1-p(t)$.
\item The probability of HIV-positive: $\Pr(T\le t | A(t)=1) = p(t)$.
\begin{itemize}
\item The probability of HIV-positive and subsequently classified as recently infected based on the recency test: $P_{rec}(t) = \Pr(M\in\calR, T\le t| A(t)=1)$.
\end{itemize}
\end{itemize}
The variable $M$ denotes the biomarker values of the HIV recency test and $\calR$ is a region for those values that classifies a subject as HIV recent.
One example of a recently proposed HIV recency test is based on the combination of three biomarkers: LAg Avidity assay, BioRad Avidity assay, and viral load.
For this test, the test-recent region $\calR$ is defined as LAg Avidity OD$_n$ $<$ 2.8, BioRad Avidity OD$_n$ $<$ 95\%, and viral load $>$ 400 copies/ml\citep{laeyendecker2018identification}.

At time $t$ when the cross-sectional sample is taken, the probability of test-recent, i.e., $M \in \mathcal{R}$, shall depend on the true infection duration $t-T$.
In particular, we define the duration-specific test-recent probability
\[\phi(u,t) = \Pr(M\in\calR|T=t-u, A(t)=1),\]
for infection duration $u\ge 0$.
Since the recency test is always applied to an HIV-positive subject that is eligible at time $t$, $\phi(u,t)$ is a probability conditional on $A(t)=1$.
We assume that $\phi(u,t)$ depends only on $u$, the infection duration, and does not depend on $t$.
That is, the calendar time when the test is taken is irrelevant to test accuracy given a fixed infection duration, and we denote the quantity as $\phi(u)$.

Summary measures of the duration-specific test-recent probability function $\phi(u)$ are suggested in literature to describe recency test properties and are used as parameters in cross-sectional incidence estimation.
For example, the mean window period ($\mu$) used in the snapshot estimator can be defined as an integration of $\phi(\cdot)$ (as long as the integration is finite) i.e., 
\[\mu = \int_0^\infty \phi(u)du.\]
The mean duration of infection (MDRI, $\Omega_{T^*}$) in the adjusted estimator is defined as a truncated integration of $\phi(\cdot)$ from $0$ to $T^*$, i.e.,
\[\Omega_{T^*} = \int_0^{T^*} \phi(u)du.\]
The false-recent rate (FRR, $\beta_{T^*}$) is defined as the probability that a randomly chosen person from the population of long-infected subjects (i.e., has an infection duration for more than time $T^*$) will be classified as “recently” infected by the recency test \citep{kassanjee2012new}.
Let $G(u)$ be the distribution of infection times among these long-infected subjects.
Then $\beta_{T^*}$ can be written as
\[\beta_{T^*} = \frac{\int_{T^*}^\infty \phi(u)dG(u)}{\int_{T^*}^\infty dG(u)}.\]

\begin{remark}
In \citet{kassanjee2012new}, the MDRI is defined as
\[\int_{0}^{T^*} P_R(u)du,\]
where $P_R (u) $ is the probability of a person infected $u$ time units ago still being alive and ``recent'' (in this setting being alive is the only eligibility criterion).
That is, 
\[P_R (u) = \Pr(A(t)=1,M\in \calR|T=t-u,A(t-u)=1).\]
This definition is different from ours in that $\phi(u)$ conditions on $A(t)=1$ but $P_R(u)$ involves the probability of $A(t)=1$ conditioning on $A(t-u)=1$.
Determining MDRI of a recency test typically involves sampling eligible individuals with known infection duration (to some reasonable approximation).
In order to be sampled at time $t$, they need to be eligible at the current time $t$ ($A(t)=1)$), instead of eligible at the time of infection ($A(t-u)=1$).
Therefore, our definition of MDRI based on $\phi(u)$ is the one that is aligned with the sampling strategy of studies that are conducted in practice.
\end{remark}

Based on our notation, the test-recent probability is given by 
\begin{align*}
P_{rec}(t) =&\int_{0}^\infty\Pr(T=t-u, M\in\calR|A(t)=1)du\\
=&\int_0^{\infty}\Pr( M\in\calR|T=t-u, A(t)=1)\Pr(T=t-u|T\le t,A(t)=1)\Pr(T\le t|A(t)=1)du\\
=&p(t)\int_0^{\infty}\phi(u) \Pr(T=t-u | T\le t, A(t)=1)du.
\end{align*}
That is, the test-recent probability is a weighted version of the duration-specific test-recent probabilities, where the weight is related to the distribution of the infection time for those infected and eligible at time $t$.
The probability $ \Pr(T=t-u | T\le t, A(t)=1)$ is not directly linked to $\lambda(t)$, the quantity of interest.
Some further assumptions are needed to construct this linkage such that the estimation of $\lambda(t)$ based on the cross-sectional sample is valid.

\subsection{Assumptions for Cross-Sectional Estimators}
Suppose that the prevalence function of HIV, $p(t)$, is continuous over time.
We introduce the following set of assumptions for cross-sectional incidence estimation.
\begin{assumpsnap}\label{ass:snapshot1}
$\phi(u)=0$ with $u$ greater than some large value.
Let $\tau$ be the upper bound of $u$ such that $\phi(u)$ is positive, i.e., $\tau = \max_u\{\phi(u)>0\}$.
\end{assumpsnap}
\begin{assumpsnap}\label{ass:snapshot2}
$\Pr(T=t-u | T\le t, A(t)=1) = \Pr(T=t| T\le t, A(t)=1)$ for all $u\in[0,\tau]$.
\end{assumpsnap}
Assumption \ref{ass:snapshot1} indicates that the tail of $\phi(u)$ goes to zero when $u$ is large, indicating zero test-recent probability for a subject infected long enough.
It would ensure that the mean window period $\mu$ is finite, which is a key requirement for the validity of the snapshot estimator.
Assumption \ref{ass:snapshot2} suggests that the infection time is uniformly distributed in $[t-\tau,t]$ for an infected eligible subject at time $t$.
Note that this is not necessarily equivalent to a constant incidence in $[t-\tau, t]$, and we will discuss this in detail in Section \ref{sec:infection-dist}.
Given Assumptions \ref{ass:snapshot1}-\ref{ass:snapshot2}, the test-recent probability can be written as
\begin{align*}
P_{rec}(t) =& p(t)\int_0^\infty\phi(u)du \Pr(T=t | T\le t, A(t)=1) = \mu \Pr(T=t | A(t)=1)\\
=&\mu \Pr(T=t |T\ge t, A(t)=1)\Pr(T\ge t)
= \mu \lambda(t)\{1-p(t-)\} = \mu \lambda(t)\{1-p(t)\},
\end{align*}
such that 
\[\lambda(t) = \frac{P_{rec}(t)}{\mu\{1-p(t)\}}.\]
By replacing the parameters with their estimators, an estimator for $\lambda(t)$ can be formulated as
\[\wlambda = \frac{N_{rec}}{N_{neg}\wmu}.\]
This estimator is indeed the snapshot estimator\citep{kaplan1999snapshot}.

Some alternative assumptions may be considered for the adjusted estimator\citep{kassanjee2012new}.
\begin{assumpkan}\label{ass:kassanjee1}
$\phi(u)$ is constant for $u\ge T^*$.
The constant value is given by $\beta_{T^*}$.
\end{assumpkan}
\begin{assumpkan}\label{ass:kassanjee2}
$\Pr(T=t-u | T\le t, A(t)=1)= \Pr(T=t| T\le t, A(t)=1)$ for all $u\in[0,T^*]$.
\end{assumpkan}
Assumption \ref{ass:kassanjee1} allows a non-zero test-recent probability for a long-infected subject, however, it restricts this probability to be constant.
Otherwise, the false-recent rate would depend on $G(\cdot)$, the distribution of infection time with respect to which the false-recent rate is evaluated.
Assumption \ref{ass:kassanjee1} may not necessarily be less restrictive than Assumption \ref{ass:snapshot1}.
For example, Assay 2A in our simulation shown in Figure \ref{fig:phi-settings} satisfies Assumption \ref{ass:snapshot1} but not Assumption \ref{ass:kassanjee1}.
Assumption \ref{ass:kassanjee2} suggests that the infection time is uniformly distributed in $[t-T^*,t]$ for an infected eligible subject at time $t$.
Since $T^*$ is usually smaller than $\tau$, Assumption \ref{ass:kassanjee2} is less restrictive than Assumption \ref{ass:snapshot2}, since the uniform distribution requirement on infection times is on a shorter time span in the past.
Given Assumptions \ref{ass:kassanjee1} and \ref{ass:kassanjee2}, the probability of test-recent can be written as
\begin{align*}
P_{rec}(t) =&p(t)\left\{\int_0^{T^*}\phi(u) \Pr(T=t-u | T\le t, A(t)=1)du+\int_{T^*}^{\infty}\beta_{T^*} \Pr(T=t-u | T\le t, A(t)=1)du\right\}\\
=&p(t)\left[\int_0^{T^*}\phi(u) \Pr(T=t-u | T\le t, A(t)=1)du+\beta_{T^*}\left\{1-\int_0^{T^*}\Pr(T=t-u | T\le t, A(t)=1)du\right\}\right]\\
=&p(t)\left[\Omega_{T^*} \Pr(T=t | T\le t,A(t)=1)+ \beta_{T^*}\left\{1-T^*\Pr(T=t |T\le t, A(t)=1)\right\}\right]\\
=&\left(\Omega_{T^*} -\beta_{T^*}T^*\right) \Pr(T=t | A(t)=1) + p(t) \beta_{T^*}\\
=&\left(\Omega_{T^*} -\beta_{T^*}T^*\right) \Pr(T=t | T\ge t, A(t)=1)\Pr(T\ge t|A(t) = 1) + p(t) \beta_{T^*}\\
=&\left(\Omega_{T^*} - \beta_{T^*}T^*\right)\lambda(t)\{1-p(t)\} + p(t)\beta_{T^*},
\end{align*}
such that
\[\lambda(t) = \frac{P_{rec}(t) - \beta_{T^*}p(t)}{\{1-p(t)\}(\Omega_{T^*} - \beta_{T^*}T^*)}.\]
Then, an estimator for $\lambda(t)$ can be formulated as
\[
\tlambda = \frac{N_{rec} - N_{pos} \wbeta_{T^*}}{N_{neg}(\wOmega_{T^*} - \wbeta_{T^*}T^*)},\]
which is the adjusted estimator\citep{kassanjee2012new}.

Assumption \ref{ass:kassanjee1} requires a constant $\phi(u)$ for $u\ge T^*$, such that the false-recent rate $\beta_T^*$ no longer depends on the distribution of the long-infected population $G(\cdot)$.
Then, an unbiased estimate of $\beta_{T^*}$ can be obtained by taking the average test-recent rate among an arbitrary sample of long-infected subjects.
In practice, $\phi(u)$ may be non-constant for $u>T^*$.
In that case, the summary FRR $\beta_{T^*}$ depends on the distribution $G(\cdot)$ and is context-specific.
For example, it may depend on the demographic and epidemiological history of the population \citep{kassanjee2016viral}.
An estimate $\wbeta_{T^*}$ depends on the distribution of long-infected subjects based on which $\wbeta_{T^*}$ is estimated.
In practice, researchers usually prefer a recency test with a small FRR ($<2\%$), so that $\phi(u)$ can be viewed as approximately constant for $u>T^*$.

\begin{remark} \label{rem:FRR}
In the case when Assumption \ref{ass:kassanjee1} is violated, use of the adjusted estimator may still be appropriate if FRR is evaluated among a similar population as the long-infected subjects in the cross-sectional sample.
Specifically, if $\Pr(T = t-u|T\le t,A(t)=1)du = dG(u)$ for $u\ge T^*$, then
\begin{align*}
& \int_{T^*}^\infty\phi(u)\Pr(t-u|T\le t,A(t) =1)du
= \int_{T^*}^\infty \phi(u)dG(u) = \beta_{T^*}\int_{T^*}^\infty dG(u) \\
=& \beta_{T^*}\int_{T^*}^\infty \Pr(T = t-u|T\le t,A(t)=1)du
= \beta_{T^*}\left\{1-\int_0^{T^*}\Pr(T=t-u|T\le t,A(t)=1)du\right\}\\
=& \beta_{T^*}\{1-T^* \Pr(T=t|T\le t,A(t)=1)\},
\end{align*}
where the last equality follows from Assumption \ref{ass:kassanjee2},.
Then, derivations for $P_{rec}(t)$ to obtain the adjusted estimator still hold.
Therefore, we may still appropriately use the adjusted estimator, if the distributions of the long-infected subjects in the cross-sectional sample and in the evaluating external study where $\wbeta_{T^*}$ is estimated are the same.
\end{remark}

\subsection{Distribution of Infection Time in the Eligible Population}\label{sec:infection-dist}
In the derivations for both estimators, one key assumption is that $\Pr(T=s | T\le t, A(t)=1) = \Pr(T=t| T\le t, A(t)=1)$ for $s\in[t-c,t]$, where $c=\tau$ for the snapshot estimator and $c=T^*$ for the adjusted estimator.
A similar assumption was also suggested in \citet{mahiane2014mixture} in describing the sensitivity and specificity of recency biomarker.

Write $\lambda_t^*(s) = \Pr(T=s |T\ge s,A(t)=1)$  and $p_t^*(s) = \Pr(T\le s|A(t)=1)$ as the incidence and prevalence at time $s$ restricted to the eligible population at time $t$.
Note that  $\lambda_t^*(s)$ differs from the $\lambda(s)$ defined in (\ref{equ:inc}), since they are restricted to the populations that are eligible at different times.
Obviously we have $\lambda_t^*(t) = \lambda(t)$ and $p_t^*(t) = p(t)$.
The key quantity $\Pr(T=s | T\le t, A(t)=1)$ can be written as
\[\Pr(T=s | T\le t, A(t)=1) =\frac{\Pr(T=s |T\ge s,A(t)=1)\Pr(T\ge s|A(t)=1)}{\Pr(T\le t |A(t)=1)} = \frac{\lambda_t^*(s)\{1-p_t^*(s-)\}}{p(t)}.\]
To connect $\lambda_t^*(s)$ and $p_t^*(s)$ with the observed incidence $\lambda(t)$ and prevalence $p(t)$, we make the following assumption.
\begin{assumpC}\label{ass:approx}
For $s\in[t-c,t]$, the restricted incidence is equal to the unrestricted (or observed) incidence, i.e., $\lambda_t^*(s) = \lambda(s)$, and the restricted prevalence is equal to the unrestricted prevalence, i.e., $p_t^*(s) = p(s)$, and for all $t$.
\end{assumpC}
This assumption would approximately hold when $c$ is small.
Specifically, if $A(t)$ is defined by characteristics such that only a small proportion of the subjects move in and out of the eligible population in a time span of $c$, then the eligible population remains approximately the same, i.e., $A(s)\approx A(t)$ for $s\in[t-c,t]$.
When A(t) is defined by covariate values such as membership of a particular population (e.g., MSM), this assumption requires that the most of the subjects who were part of this population at time $s$ are also part of this population at time $t$.
Based on Assumption \ref{ass:approx}, 
\[\Pr(T=s | T\le t, A(t)=1) = \frac{\lambda(s)\{1-p(s)\}}{p(t)}\]
for $s\in[t-c,t]$.
Therefore, Assumptions \ref{ass:snapshot2} and \ref{ass:kassanjee2} would hold if the incidence and prevalence are constant over $[t-c,t]$ ($c=\tau$ for the snapshot estimator and $c=T^*$ for the adjusted estimator).
Specifically, we consider the following assumption.
\begin{assumpD}\label{ass:const}
$\lambda(s) = \lambda(t)$ and $p(s) = p(t)$ for $s\in[t-c,t]$, and for all $t$.
\end{assumpD}

To summarize these assumptions, the consistency of the snapshot estimator and adjusted estimator is given in the following theorems.

\begin{theorem}\label{thm:snapshot}
Suppose that Assumptions \ref{ass:approx} and \ref{ass:const} hold for $c=\tau$.
Then, Assumption \ref{ass:snapshot2} holds.
If we further assume Assumption \ref{ass:snapshot1}, then the snapshot estimator $\wlambda$ is unbiased for estimating $\lambda(t)$.
\end{theorem}

\begin{theorem}\label{thm:kassanjee}
Suppose that Assumptions \ref{ass:approx} and \ref{ass:const} hold for $c=T^*$.
Then, Assumption \ref{ass:kassanjee2} holds.
If we further assume Assumption  \ref{ass:kassanjee1}, then the adjusted estimator $\tlambda$ is unbiased for estimating $\lambda(t)$.
\end{theorem}

\subsection{Violation of Constant Incidence and Prevalence}\label{expected-bias}
We have given results on consistency of the snapshot and adjusted estimators with Theorems \ref{thm:snapshot} and \ref{thm:kassanjee}.
The main epidemiological requirement is Assumption \ref{ass:const}, i.e., incidence and prevalence are constant over a period of time.
In this section, we explore the expected bias when Assumption \ref{ass:const} fails to hold (but all other assumptions hold).
Specifically, we assess the bias associated with non-constant incidence $\lambda(t)$ but constant prevalence.
By Assumption \ref{ass:approx}, 
\begin{align*}
    P_{rec}(t) = (1-p)\int_0^\infty\phi(u)\lambda(t-u)du,
\end{align*}
where $p$ is the constant prevalence.

We first consider the bias of the snapshot estimator.
Given Assumption \ref{ass:snapshot1}, the expected value of the snapshot estimator is given by 
\[E(\wlambda) = \int_0^\tau\frac{\phi(u)}{\mu}\lambda(t-u)du,\]
which is a weighted version of the incidence over $[t-\tau,t]$.
With an HIV epidemic 
If the incidence $\lambda(s)$ is linearly changing in $s\in[t-\tau,t]$, i.e., $\lambda(s) = \lambda(t) +\rho(t-s)$, then 
\begin{align*}
    E(\wlambda) = \int_0^\tau\frac{\phi(u)}{\mu}\{\lambda(t)+\rho u\} du = \lambda(t) + \rho\frac{\int_0^\tau u\phi(u)du}{\mu} =\lambda(t-\omega),
\end{align*}
where 
$\omega = \int_0^\tau u\phi(u)du /\mu$ is the mean shadow time defined by Kaplan and Brookmeyer \citep{kaplan1999snapshot}, indicating that the cross-sectional sample is ``casting a shadow'' back in time.
That is, when the incidence is linearly changing in time and the prevalence is constant, the snapshot estimator estimates the incidence rate $\omega$ time units ago.
The estimation bias is given by $E(\wlambda) -\lambda(t) = \lambda(t-\omega) - \lambda(t) = \rho\omega$.
For example, if incidence is decreasing, i.e., $\rho>0$, the underlying incidence that produced an infection $u>0$ time units ago was higher than the current incidence, so the estimate $\hat{\lambda}$ will have positive bias.

Similarly, for the adjusted estimator, we evaluate the expected value under Assumption \ref{ass:kassanjee1}, which is given by
\begin{align*}
    E(\tlambda) = \int_0^{T^*}\frac{\phi(u)-\beta_{T^*}}{\Omega_{T^*} - T^*\beta_{T^*}}\lambda(t-u)du.
\end{align*}
When $\lambda(s) = \lambda(t) + \rho(t-s)$ for $s\in[t-T^*,t]$,
\begin{align*}
    E(\tlambda) = \lambda(t) + \rho\frac{\int_0^{T^*} u\{\phi(u)-\beta_{T^*}\}du}{\Omega_{T^*} -T^* \beta_{T^*}} =\lambda(t-\omega^*),
\end{align*}
where $\omega^* = \int_0^{T^*} u\{\phi(u)-\beta_{T^*}\}du/ (\Omega_{T^*} - T^* \beta_{T^*})$.
It can be viewed as a ``mean shadow time'' for the adjusted estimator with a recency test that satisfies Assumption \ref{ass:kassanjee1}. 
The estimation bias is given by $E(\tlambda) -\lambda(t) = \lambda(t-\omega^*) - \lambda(t) = \rho\omega^*$.

Thus far, we have provided a framework with precisely defined assumptions through which to understand both the snapshot and adjusted estimators.
To our knowledge, rigorous derivation of the estimators and their assumptions has not been done by others.
In the following section, we evaluate how these estimators perform empirically under realistic epidemiological scenarios and with realistic recency test algorithms.

\section{Simulations}\label{simulations}
To evaluate the numerical performance of the estimators under various settings, we conducted simulation studies.
Throughout the simulations, we assume that Assumption \ref{ass:approx} on approximation of the eligible population always hold.
We also assume that prevalence is constant over time.
We consider different settings of HIV epidemics and recency tests, where Assumptions \ref{ass:snapshot1}, \ref{ass:kassanjee1}, and \ref{ass:const} hold or not.
We calculate the snapshot and adjusted estimators using (\ref{equ:snapshot}) and (\ref{equ:kassanjee}), with variance estimators calculated based on Appendix A of \citet{gao2020sample}.
Importantly, these variance estimators accounts for variability in estimating $\mu$, $\Omega_{T^*}$ and $\beta_{T^*}$ from an external study.

\subsection{Epidemiological Parameters}\label{sim:epi}
We generate practical settings by mimicking the epidemiological dynamics of HIV in a population of men who have sex with men (MSM) attending Silom Community Clinic in Bangkok, Thailand \citep{pattanasin2020recent}.
Particularly, we set the prevalence to be constant and as the mean prevalence in 2011-2018 in that population, and generate settings with different incidences by modeling the HIV incidence in 2011-2018 by either a linear model or log-linear model, to reflect a linearly decreasing or exponentially decreasing incidence.
Based on the estimates from the Bangkok MSM data, we consider the following settings corresponding to different trends in HIV incidence.
\begin{enumerate}[label=(\alph*)]
    \item Constant incidence: $\lambda(s) = 0.032$, $p = 0.29$.
    \item Linearly decreasing incidence: $\lambda(s) = 0.032 + 0.0028(t-s)$, $p=0.29$.
    \item Exponentially decreasing incidence: $\lambda(s) = 0.032 \exp\{0.07(t-s)\} $, $p=0.29$.
\end{enumerate}

Assumption \ref{ass:const} is satisfied when the incidence is constant, and it is violated in the linear and exponential settings.
We would like to estimate the incidence at time $t$ such that the ``true'' incidence is 0.032 across all settings.
In our simulations, we will assess how violating this assumption affects the performance of the snapshot and adjusted estimators.

\subsection{Recency Test Characteristics}
We sought to assess the performance of the two estimators with a variety of recency tests with different characteristics.
The properties of those simulated recency tests mimic two tests in \citet{brookmeyer2013estimation} and \citet{laeyendecker2018identification}, with modifications that allow us to assess the performance of the estimators under diverse conditions.
For the snapshot estimator, we set $\tau = 12$.
For the adjusted estimator, we always considered $T^* = 2$, i.e., any person with an infection acquired longer than 2 years ago is a ``long-infected'' case.

We first consider a set of recency tests with a relatively short mean window period and a short shadow period, mimicking a recency test in \citet{brookmeyer2013estimation} that classifies a subject as recenct if their BED capture enzyme immunoassay (BED-CEIA) $\le$ 1.5, their Bio-Rad Avidity (BRAI) (Bio-Rad Laboratories, Mississauga, ON) $<$ 40, and their viral load $>$ 400 copies/ml.
This test has a mean window period of 101 days and a shadow period of 194 days.
We generated four different recency tests that mimic this test:

\begin{enumerate}[label=(1\Alph*)]
\item $\phi_{1A}(t) = 1 - F_{Gamma}(t; \alpha=0.352, \beta=1.273)$, where $F_{Gamma}(\cdot;\alpha,\beta)$ is the cumulative distribution function of a Gamma random variable with shape $\alpha$ and rate $\beta$.
Assumption \ref{ass:snapshot1} (approximately) holds for this test with mean window period 101 days and mean shadow 194 days.
Assumption \ref{ass:kassanjee1} fails to hold since $\phi_{1A}(t)$ is non-constant for $t\ge 2$.
MDRI = 98 days and the test-recent rate probability at $t=2$ is 1.4\%.
\item $\phi_{1B}(t) = \phi_{1A}(t) I(t \leq 2) + 0.014 I(t > 2)$.
This test modifies test 1A by carrying forward the 1.4\% test-recent probability at $t=2$, such that Assumption \ref{ass:kassanjee1} for the adjusted estimator holds with MDRI = 98 days and FRR = 1.4\%.
Assumption \ref{ass:snapshot1} for the snapshot estimator no longer holds such that the mean window period is infinite.
\item $\phi_{1C}(t) = \phi_{1B}(t) + f_N(t; 7, 1) / 8$, where $f_N(t; \mu, \sigma)$ is the density function of a normal random variable with mean $\mu$ and standard deviation $\sigma$.
This test further modifies test 1B by adding a normally distributed spike centered at 7 years, such that Assumption \ref{ass:kassanjee1} on constant false-recent rate no longer holds.
This test, similar to that depicted in the figure of epidemiological and test recent dynamics in \citet{kassanjee2012new}, represents a setting in which individuals who have been on antiretroviral therapy for years may have biomarker profiles similar to those who have been recently infected, and thus the false-recent rate among those individuals is relatively higher.
\item $\phi_{1D}(t) = \phi_{1B}(t) + F_{N}(t; 10, 2) / 10$ where $F_{N}(t; \mu, \sigma)$ is the cumulative distribution function of a normal random variable with mean $\mu$ and standard deviation $\sigma$.
This test modifies test 1B by steadily increasing the false-recent rate starting around 6 years, and reaches 9.8\% at 12 years.
The high false-recent rate is motivated by the BED assay\citep{parekh2002quantitative}, which has been shown to have an FRR in some populations up to 15\%\citep{mastro2010estimating}.
\end{enumerate}

\begin{figure}[htb!]
\centering
\includegraphics[width=6in]{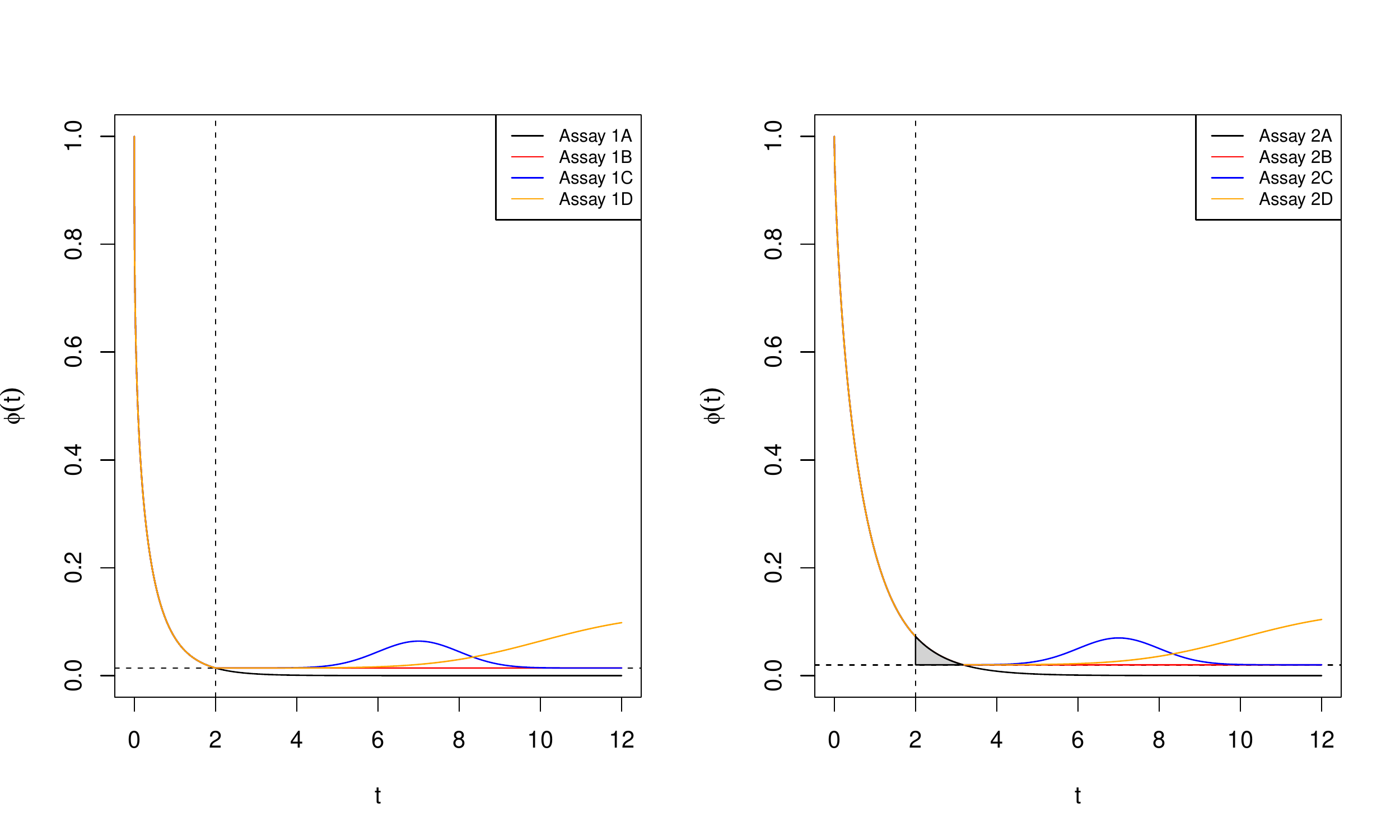}
\caption{Plot of duration-specific test-recent probability $\phi(t)$ for eight recency tests.
Left: Tests 1A-D.
Right: Tests 2A-D.}\label{fig:phi-settings}
\end{figure}

Additionally, we consider another set of recency tests with a longer mean window period and a longer shadow period.
This set of tests was modeled after the one presented in \citet{laeyendecker2018identification} for HIV sub-type C, LAg Avidity (Sedia HIV-1 LAg Avidity EIA; Sedia Biosciences Corporation, Portland, OR, USA) ODn $\le$ 2.8 BRAI $\le$ 95, and viral load $>$ 400 copies/ml.
It has a mean window period of 248 days and a shadow period of 306 days.
We generated four recency tests that mimic this test:

\begin{enumerate}[label=(2\Alph*)]
    \item $\phi_{2A}(t) = 1 - F_{Gamma}(t; \alpha=0.681, \beta=1.003)$.
    Assumption \ref{ass:snapshot1} holds for this test with mean window period 248 days and shadow 306 days.
    Assumption \ref{ass:kassanjee1} fails to hold since $\phi_{2A}(t)$ is non-constant for $t\ge 2$.
    MDRI = 224 days and the test-recent rate probability at $t=2$ is 7.25\%.
    \item $\phi_{2B}(t) = \phi_{2A}(t) I(t \leq 3.17) + 0.020 I(t > 3.17)$.
    This recency test has a constant 2\% test-recent probability when $t\ge 3.17$.
    Unlike test 1B, Assumption \ref{ass:kassanjee1} for the adjusted estimator is violated since the test-recent rate is non-constant after year $2$, as depicted in Figure \ref{fig:phi-settings} by the shaded grey region.
    \item $\phi_{2C}(t) = \phi_{2B}(t) + f_N(t; 7, 1) / 8$.
    A similar normally distributed spike centered at 7 years was added to test 2B.
    \item $\phi_{2D}(t) = \phi_{2B}(t) + F_N(t; 10, 2) / 10$.
    Similar to $\phi_{1D}$, FRR increases up to 10.4\% at time 12.
\end{enumerate}
The duration-specific test-recent probabilities of all six recency tests are depicted in Figure \ref{fig:phi-settings}.
The shaded grey area highlighted that Assumption \ref{ass:kassanjee1} is violated for test 2B, even though it holds for test 1B: there is a non-constant fraction of the long-infected subjects (infection duration $> 2$) who test recent.

\subsection{Data Simulation Procedure}

The data simulation procedure consists of two parts.
The first part requires simulating data to mimic an external study based on which we estimate the properties of a particular recency test, including mean window period, MDRI and FRR.
The second part includes simulation of cross-sectional samples from a population with given HIV epidemiological dynamics.
These separate data simulation procedures are outlined in sections \ref{sim:external} and \ref{sim:xsectional}, respectively.

\subsubsection{External Study and Estimation of Recency Assay Parameters}\label{sim:external}

Here we outline the process of simulating recency test results for samples (with known infection durations) in an external study and estimating recency test parameters based on such simulated data.
We simulate infection durations in the external study similar to those in \citet{duong2015recalibration}, with detailed procedure described in Appendix \ref{append:estphi}.
Then, given a duration-specific test-positive probability $\phi(\cdot)$, we generate a test-recent indicator  by $\Delta_{ij} \sim \mathrm{Bernoulli}(\phi(u_{ij}))$, where $u_{ij}$ is the simulated infection duration for sample $j=1,\dots, n_i$ of subject $i=1,\dots,m$ in the external study.
Based on the observed data $\{(u_{ij},\Delta_{ij}):i=1,\dots,n;j=1,\dots,n_i\}$ in the external study, we estimate the function $\phi(t)$ using generalized estimating equations \citep{liang1986gee}, with  an exchangeable correlation structure accounting for within-subject correlation.
The marginal model uses a logit link and a cubic polynomial for the linear predictor by assuming
\begin{align*}
    \mathrm{logit}(E[\Delta_{ij}]) = \gamma_0 + \gamma_1 u_{ij} + \gamma_2 u_{ij}^2 + \gamma_3 u_{ij}^3.
\end{align*}
Then, an estimate for $\phi(u)$ can be constructed by $\wphi(u) = \wgamma_0 + \wgamma_1 u + \wgamma_2 u^2 + \wgamma_3 u^3$, where $\wbgamma = (\wgamma_0, \wgamma_1, \wgamma_2, \wgamma_3)$ is the parameter estimate.
We use robust standard errors for variance estimation.
We then calculate the mean window period and MDRI by numerically integrating the estimated the recency test-positive function $\wphi$.
In particular, the mean window period involves a numerically integrating until infinity, however, in the simulation we set the upper bound of the integration to the maximum duration observed in the simulated sample (approximately 8 years).
We estimate the variance of $\wOmega_{T^*}$ and $\wmu$ using the delta method and the robust variance-covariance matrix of $\wbgamma$.
We estimate FRR by evaluating the average test-recent probability among a number of long-infected subjects.
In particular, we consider 1500 long-infected subjects with duration of infection uniformly distributed between $T^*=2$ and $\tau=12$ years, similar to other studies\citep{kassanjee2016viral}.

\subsubsection{Simulation of Cross-Sectional Samples}\label{sim:xsectional}

To generate a cross-sectional sample with fixed size $N$, we first simulate the number of HIV positive subjects $N_{pos} \sim \mathrm{Binomial}(N, p)$, where $p$ is the (constant) prevalence.
The number of HIV negative subjects is given by $N_{neg} = N - N_{pos}$.
For each HIV-positive subject $i=1,\dots, N_{pos}$, we simulate their infection duration based on the epidemic parameters $(p,\lambda(t))$, with details given in Appendix \ref{append:simudist}.
Given $T_i$, we generate a recency test indicator $\Delta_i\sim \mathrm{Bernoulli}(\phi(t-T_i))$.
Finally, we calculate $N_{rec} = \sum_{i=1}^{N_{pos}}\Delta_i$.

\subsubsection{Incidence Estimation}
For each simulation replicate, we first generate observations of an external study to obtain the estimates $\wmu$, $\wOmega_{T^*}$ and $\wbeta_{T^*}$.
Then, we generate an independent cross-sectional sample $\{N_{rec}, N_{pos},N_{neg}\}$.
We calculate $\wlambda$ by Equation (\ref{equ:snapshot}) and calculate $\tlambda(t)$ by Equation (\ref{equ:kassanjee}).
We estimate their variances  based on the formulas in appendix A of \citet{gao2020sample}.

\subsection{Software}
The code to reproduce these simulations, and instructions to use functions for estimating incidence based on cross-sectional data are available.\footnote{\url{https://github.com/mbannick/XSRecency/tree/0.0.0}}

\subsection{Results}
Table \ref{tab:sim-results} shows the simulation results in settings with constant, linear, and exponential incidence trends and recency tests 1A-D and 2A-D, with fixed cross-sectional trial sample size $N = 5000$.
Across all settings, the ``true'' value for incidence is $\lambda = 0.032$.
Each entry is based on 5,000 simulations.

\begin{table}[htbp]
\centering
\begin{tabular}{cccccccccccc}
  \hline
  \multicolumn{2}{c}{Setting} & \multicolumn{5}{c}{Snapshot Estimator (\ref{equ:snapshot})} & \multicolumn{5}{c}{Adjusted Estimator (\ref{equ:kassanjee})} \\
 \hline  Incidence & Assay  & Asm. & Bias & SE & SEE & Cov & Asm. & Bias & SE & SEE & Cov \\
 \hline  \hline  \multicolumn{11}{c}{Recency Assay 1A-D} \\
 \hline
Constant & 1A & \checkmark & 0.04 & 0.63 & 0.63 & 94.54 & $\times$ & 0.05 & 0.67 & 0.66 & 94.52 \\ 
  Linear & 1A & $\times$ & 0.21 & 0.64 & 0.64 & 95.24 & $\times$ & 0.22 & 0.68 & 0.68 & 95.00 \\ 
  Exponential & 1A & $\times$ & 0.18 & 0.64 & 0.64 & 95.32 & $\times$ & 0.18 & 0.68 & 0.67 & 95.00 \\ 
   \hline Constant & 1B & $\times$ & 0.79 & 0.68 & 0.68 & 84.24 & \checkmark & 0.11 & 0.99 & 0.99 & 95.40 \\ 
  Linear & 1B & $\times$ & 0.87 & 0.69 & 0.69 & 81.42 & $\times$ & 0.23 & 1.00 & 1.00 & 95.02 \\ 
  Exponential & 1B & $\times$ & 0.85 & 0.69 & 0.69 & 82.06 & $\times$ & 0.20 & 1.00 & 1.00 & 95.08 \\ 
   \hline Constant & 1C & $\times$ & 0.73 & 0.71 & 0.69 & 84.54 & $\times$ & -0.01 & 1.32 & 1.32 & 95.58 \\ 
  Linear & 1C & $\times$ & 1.32 & 0.80 & 0.76 & 60.66 & $\times$ & 1.25 & 1.37 & 1.38 & 86.28 \\ 
  Exponential & 1C & $\times$ & 1.33 & 0.80 & 0.77 & 60.24 & $\times$ & 1.28 & 1.38 & 1.38 & 85.70 \\ 
     \hline Constant & 1D & $\times$ & 3.28 & 0.98 & 0.97 & 03.86 & $\times$ & 1.01 & 1.62 & 1.63 & 90.66 \\ 
  Linear & 1D & $\times$ & 1.41 & 0.77 & 0.77 & 54.56 & $\times$ & -2.45 & 1.55 & 1.54 & 64.23 \\ 
  Exponential & 1D & $\times$ & 1.41 & 0.77 & 0.77 & 54.32 & $\times$ & -2.44 & 1.55 & 1.54 & 64.59 \\ 
   \hline  \multicolumn{11}{c}{Recency Assay 2A-D} \\
 \hline Constant & 2A & \checkmark & 0.06 & 0.40 & 0.40 & 95.10 & $\times$ & 0.05 & 0.46 & 0.46 & 95.24 \\ 
  Linear & 2A & $\times$ & 0.29 & 0.41 & 0.42 & 92.44 & $\times$ & 0.33 & 0.48 & 0.48 & 91.72 \\ 
  Exponential & 2A & $\times$ & 0.26 & 0.41 & 0.42 & 93.40 & $\times$ & 0.28 & 0.47 & 0.48 & 92.70 \\ 
   \hline Constant & 2B & $\times$ & 0.48 & 0.43 & 0.43 & 83.74 & $\times$ & 0.08 & 0.57 & 0.57 & 94.96 \\ 
  Linear & 2B & $\times$ & 0.62 & 0.44 & 0.44 & 74.06 & $\times$ & 0.28 & 0.57 & 0.58 & 93.18 \\ 
  Exponential & 2B & $\times$ & 0.59 & 0.44 & 0.44 & 75.74 & $\times$ & 0.25 & 0.57 & 0.58 & 93.78 \\ 
   \hline Constant & 2C & $\times$ & 0.43 & 0.46 & 0.44 & 84.76 & $\times$ & 0.03 & 0.65 & 0.65 & 95.46 \\ 
  Linear & 2C & $\times$ & 0.84 & 0.49 & 0.48 & 58.22 & $\times$ & 0.69 & 0.67 & 0.68 & 84.98 \\ 
  Exponential & 2C & $\times$ & 0.83 & 0.49 & 0.48 & 59.34 & $\times$ & 0.67 & 0.67 & 0.68 & 85.54 \\ 
  \hline Constant & 2D & $\times$ & 1.62 & 0.52 & 0.52 & 09.06 & $\times$ & 0.40 & 0.72 & 0.73 & 92.08 \\ 
  Linear & 2D & $\times$ & 0.88 & 0.46 & 0.47 & 53.68 & $\times$ & -0.68 & 0.69 & 0.69 & 82.04 \\ 
  Exponential & 2D & $\times$ & 0.87 & 0.46 & 0.47 & 54.52 & $\times$ & -0.71 & 0.69 & 0.69 & 81.18 \\ 

   \hline
\end{tabular}
\caption{Summary statistics ($\times 10^{-2}$) for the simulation studies with different settings over 5000 simulations each.
For each epidemiological setting and recency test, we show the empirical median bias (Bias), the empirical standard error (SE), the average standard error estimate (SEE), and the empirical coverage probability of the 95\% confidence intervals (Cov).
We note whether the assumptions are satisfied for the snapshot and adjusted estimator in the Asm. column.}\label{tab:sim-results}
\end{table}

\subsubsection{Snapshot Estimator}
Recency assays 1A and 2A satisfy Assumption \ref{ass:snapshot1} for the snapshot estimator.
In the constant incidence setting where Assumption \ref{ass:const} further holds, the empirical bias is small.
In the settings where Assumption \ref{ass:const} fails to hold (linear or exponential incidence), there is empirical bias associated with the snapshot estimator, and the empirical bias is close to the expected bias calculated based on the formula in Section \ref{expected-bias} (0.15$\times 10^{-2}$ and 0.12$\times 10^{-2}$ for assay 1A in the linear and exponential settings, respectively; 0.20$\times 10^{-2}$ and 0.23$\times 10^{-2}$ for assay 2A in the linear and exponential settings, respectively).
For the recency tests that violate Assumption \ref{ass:snapshot1} (assays 1B-D, 2B-D), the empirical bias is larger, and it increases if the constant incidence assumption is also violated.

Across all settings, the standard error estimate is close to the empirical standard error, indicating reasonable estimation of variability.
Snapshot estimates based on recency tests 2A-D have smaller variability than those based on 1A-D, corresponding to a larger area under the $\phi(t)$ curve.
The coverage probabilities are close to the nominal level with recency test 1A and 2A when incidence is constant.
There is under-coverage when the incidence is non-constant or when Assumption \ref{ass:snapshot1} fails to hold (recency tests 1B-D, 2B-D).

\subsubsection{Adjusted Estimator}
When incidence is constant, the empirical bias for recency tests 1A-C and 2A-C is small, while the empirical bias for recency tests 1D and 2D are relatively large.
When the constant incidence assumption is violated, the empirical bias gets larger and the empirical bias for assay 1B matches with the theoretical calculated expected bias when Assumption \ref{ass:kassanjee1} holds (0.10$\times 10^{-2}$ and 0.09$\times 10^{-2}$ for assay 1B for the linear and exponential settings, respectively).

Across all settings, the standard error estimate is close to the empirical standard error, indicating reasonable estimation of variability.
Notably, the empirical standard errors for the adjusted estimator are always larger than those for the snapshot estimator.

The coverage probabilities are close to the nominal level with recency tests 1A and 1B, even in non-constant incidence settings.
The coverage probabilities in the constant incidence setting are always close to the nominal level.
Similar to the snapshot estimator, there is under-coverage when the incidence is non-constant for some recency tests.
The coverage probabilities are closer to the nominal level compared to the snapshot estimators in most non-constant incidence settings, mainly due to a larger associated variability.

\subsection{Sensitivity to Distribution of Long-infected Subjects in External Study}
Table \ref{tab:sim-results} shows that recency tests 1C and 2C give minimal bias and nominal coverage in the constant incidence setting.
It seems to suggest that the adjusted estimator may sometimes be robust to violation of the recency test requirement (Assumption \ref{ass:kassanjee1}) when the incidence in the population is constant.
However, it is the result of a specific choice of distribution of infection duration in the external study, based on which we estimate $\beta_{T^*}$.
The infection times of the long-infected subjects were generated uniformly from 2 to 12 years, such that the estimated $\wbeta_{T^*}$ reflects the average false-recent rate in [2,12], similar to the average false-recent rate in a constant incidence setting.

To evaluate sensitivity to this distributional assumption, we consider two different distributions of infection durations for long-infected subjects in the external study in estimating $\beta_{T^*}$.
The first distribution is similar to those who were infected more than two years ago in \citet{duong2015recalibration}, where the range of infection duration is [2,8.25] years and the distribution is certainly not uniform (see Figure \ref{fig:duong}).
The second distribution is to truncate Duong's infection durations at 5 years, such that the range of infection duration is [2,5] years.

Simulation results are shown in Table \ref{tab:sens-1}.
Different sampling schemes for the long-infected subjects provide different estimates for the incidence with recency tests 1C and 2C.
In particular, if the distribution of long-infected subjects fails to recover the major trend of the $\phi$ function (e.g., the spike at 7 years for assays 1C and 2C), the estimator is biased and the coverage is poor.
This suggests caution in using the adjusted estimator when Assumption \ref{ass:kassanjee1} fails to hold, and that one should be sure that the distribution of long-infected subjects matches with the cross-sectional sample.

\begin{table}[ht]
\centering
\begin{tabular}{cccccc}
  \hline
  \multicolumn{2}{c}{Setting} & \multicolumn{4}{c}{Adjusted Estimator (\ref{equ:kassanjee})} \\
 \hline  Assay & Long-Infected Distribution & Bias & SE & SEE & Cov \\
 \hline  \hline
1C & Uniform [2,12] & -0.01 & 1.32 & 1.32 & 95.58 \\ 
  1C & \citet{duong2015recalibration} [2,8.25] & 0.56 & 1.49 & 1.26 & 88.26 \\ 
  1C & \citet{duong2015recalibration} [2,5] & 1.78 & 1.32 & 1.13 & 62.98 \\ 
  \hline
  2C & Uniform [2,12] & 0.03 & 0.65 & 0.65 & 95.46 \\ 
  2C & \citet{duong2015recalibration} [2,8.25] & 0.01 & 0.76 & 0.66 & 90.92 \\ 
  2C & \citet{duong2015recalibration} [2,5] & 0.36 & 0.77 & 0.64 & 86.82 \\ 
   \hline
\end{tabular}
\caption{Sensitivity analysis for the adjusted estimator over 5000 simulations for different sampling strategy to estimate $\beta_{T^*}$.
All samples include 1500 subjects (we re-sampled the infection duration in Duong dataset to get 1500 subjects).
Results shown for recency tests 1C and 2C.}\label{tab:sens-1}
\end{table}

\section{Discussion}

In this manuscript, we considered a unified statistical framework to assess the assumptions for cross-sectional incidence estimation based on the snapshot and Kassanjee's adjusted estimators.
We established two key assumptions: the incidence and prevalence in the population of interest are constant over the period of time preceding a cross-sectional sample; and the duration-specific test-recent probability function $\phi(t)$ goes to zero for the snapshot estimator or is constant in the tail for the adjusted estimator.
We derived the theoretical biases of the estimators when constant incidence assumption fails to hold.
To empirically assess the biases, we conducted simulation studies under various scenarios with different epidemiological settings and different recency test properties.

Indeed, the estimators perform well when their corresponding assumptions hold.
When the constant incidence assumption is violated, the numerical bias is commensurate with the theoretically calculated bias.
The adjusted estimator is more robust when the assumptions about the recency test properties (Assumption \ref{ass:snapshot1} or \ref{ass:kassanjee1}) are violated; though to compensate for this, the variability of the adjusted estimator is always larger than that of the snapshot estimator.
This robustness to mis-specification makes the adjusted estimator more flexible in the setting where the property of a specific recency test is not precisely known.

There are important differences between the snapshot and adjusted estimator with respect to their requirements.
The snapshot estimator requires a finite positive range for $\phi(t)$ (Assumption \ref{ass:snapshot1}).
In other words, if someone was infected sufficiently long time ago, the recency test is perfectly specific.
In contrast, the adjusted estimator requires a constant $\phi(t)$ in the tail (Assumption \ref{ass:kassanjee1}).
In other words, past a certain point, the false-recent test probability is unrelated to infection duration.
As illustrated by assay 2A in the simulation studies, Assumption \ref{ass:kassanjee1} is not necessarily less restrictive than Assumption \ref{ass:snapshot1} and indeed the snapshot estimator performs better for this specific assay.
In practice, to obtain the best performance, the researcher should be cautious and understand the properties of the recency test before choosing whether to apply either of the estimators.

An additional consideration when using the adjusted estimator is that the performance of the adjusted estimator may be affected by the distribution of the long-infected subjects that are used to estimate FRR.
As suggested in Remark \ref{rem:FRR}, the bias from the adjusted estimator may be minimal if the distributions of the long-infected subjects in the cross-sectional sample and in the evaluating external study are the same.
However, if the range of the infection duration of the long-infected subjects fails to recover a non-constant region of the $\phi$ function of the recency test, or if the distribution of infection duration differs much from that of long-infected subjects in the cross-sectional sample, biases from estimating FRR lead to bias and under-coverage in the adjusted estimator.

In order to use the adjusted estimator, researchers need to specify a fixed $T^*$ beyond which point subjects are regarded as long-infected.
In practice, $T^*$ is set at 1 or 2 years, and a proper test-recent region $\calR$ is then chosen to yield a recency test with the desired properties (e.g., large MDRI, FRR<2.0\%) defined upon this choice of $T^*$.
Choice of $T^*$ would affect the performance of the adjusted estimator through its impact on MDRI/FRR and the fact that the adjusted estimator estimates a weighted average incidence in a range of $T^*$ prior to the cross-sectional time.
In this manuscript, we did not assess the impact of $T^*$, since it would involve extensive modeling for the biomarker values at different infection duration.
We wish to explore the effect of different choices of $T^*$ in future research.

Finally, the performance of the adjusted estimator is sensitive to Assumption \ref{ass:kassanjee1} that includes a constant tail of the duration-specific test-recent probability $\phi(t)$, which is affected by the test-recent region $\calR$ of the recency test.
In particular, the test-recent region $\calR$ is usually chosen to guarantee such assumption, leading to a potentially small MDRI and suboptimal power for the adjusted estimator.
An alternative strategy is to directly model the infection duration and construct an incidence estimator based on a predicted infection duration given recency assay readings.
Since it uses the recency assay readings, such strategy is similar to making use of ``recency'' results from multiple test-regions so that more power may be gained.
This alternative strategy is currently under investigation.

\newpage
\bibliography{cross}

\begin{thebibliography}{}

\bibitem[\protect\citeauthoryear{Brookmeyer}{Brookmeyer}{2010}]{brookmeyer2010measuring}
Brookmeyer, R. (2010).
\newblock Measuring the hiv/aids epidemic: approaches and challenges.
\newblock {\em Epidemiologic reviews\/}~{\em 32\/}(1), 26--37.

\bibitem[\protect\citeauthoryear{Brookmeyer, Konikoff, Laeyendecker, and
  Eshleman}{Brookmeyer et~al.}{2013}]{brookmeyer2013estimation}
Brookmeyer, R., J.~Konikoff, O.~Laeyendecker, and S.~H. Eshleman (2013).
\newblock Estimation of hiv incidence using multiple biomarkers.
\newblock {\em American journal of epidemiology\/}~{\em 177\/}(3), 264--272.

\bibitem[\protect\citeauthoryear{Brookmeyer and Quinn}{Brookmeyer and
  Quinn}{1995}]{brookmeyer1995estimation}
Brookmeyer, R. and T.~C. Quinn (1995).
\newblock Estimation of current human immunodeficiency virus incidence rates
  from a cross-sectional survey using early diagnostic tests.
\newblock {\em American journal of epidemiology\/}~{\em 141\/}(2), 166--172.

\bibitem[\protect\citeauthoryear{Cousins, Laeyendecker, Beauchamp, Brookmeyer,
  Towler, Hudelson, Khaki, Koblin, Chesney, Moore, et~al.}{Cousins
  et~al.}{2011}]{cousins2011use}
Cousins, M.~M., O.~Laeyendecker, G.~Beauchamp, R.~Brookmeyer, W.~I. Towler,
  S.~E. Hudelson, L.~Khaki, B.~Koblin, M.~Chesney, R.~D. Moore, et~al. (2011).
\newblock Use of a high resolution melting (hrm) assay to compare gag, pol, and
  env diversity in adults with different stages of hiv infection.
\newblock {\em PLoS One\/}~{\em 6\/}(11), e27211.

\bibitem[\protect\citeauthoryear{Duong, Kassanjee, Welte, Morgan, De, Dobbs,
  Rottinghaus, Nkengasong, Curlin, Kittinunvorakoon, et~al.}{Duong
  et~al.}{2015}]{duong2015recalibration}
Duong, Y.~T., R.~Kassanjee, A.~Welte, M.~Morgan, A.~De, T.~Dobbs,
  E.~Rottinghaus, J.~Nkengasong, M.~E. Curlin, C.~Kittinunvorakoon, et~al.
  (2015).
\newblock Recalibration of the limiting antigen avidity eia to determine mean
  duration of recent infection in divergent hiv-1 subtypes.
\newblock {\em PloS one\/}~{\em 10\/}(2), e33328.

\bibitem[\protect\citeauthoryear{Eshleman, Hughes, Laeyendecker, Wang,
  Brookmeyer, Johnson-Lewis, Mullis, Hackett~Jr, Vallari, Justman,
  et~al.}{Eshleman et~al.}{2013}]{eshleman2013use}
Eshleman, S.~H., J.~P. Hughes, O.~Laeyendecker, J.~Wang, R.~Brookmeyer,
  L.~Johnson-Lewis, C.~E. Mullis, J.~Hackett~Jr, A.~S. Vallari, J.~Justman,
  et~al. (2013).
\newblock Use of a multifaceted approach to analyze hiv incidence in a cohort
  study of women in the united states: Hiv prevention trials network 064 study.
\newblock {\em The Journal of infectious diseases\/}~{\em 207\/}(2), 223--231.

\bibitem[\protect\citeauthoryear{Gao, Glidden, Hughes, and Donnell}{Gao
  et~al.}{2020}]{gao2020sample}
Gao, F., D.~V. Glidden, J.~P. Hughes, and D.~Donnell (2020).
\newblock Sample size calculation for active-arm trial with counterfactual
  incidence based on recency assay.

\bibitem[\protect\citeauthoryear{Hargrove, Humphrey, Mutasa, Parekh, McDougal,
  Ntozini, Chidawanyika, Moulton, Ward, Nathoo, et~al.}{Hargrove
  et~al.}{2008}]{hargrove2008improved}
Hargrove, J.~W., J.~H. Humphrey, K.~Mutasa, B.~S. Parekh, J.~S. McDougal,
  R.~Ntozini, H.~Chidawanyika, L.~H. Moulton, B.~Ward, K.~Nathoo, et~al.
  (2008).
\newblock Improved hiv-1 incidence estimates using the bed capture enzyme
  immunoassay.
\newblock {\em Aids\/}~{\em 22\/}(4), 511--518.

\bibitem[\protect\citeauthoryear{Janssen, Satten, Stramer, Rawal, O'Brien,
  Weiblen, Hecht, Jack, Cleghorn, Kahn, et~al.}{Janssen
  et~al.}{1998}]{janssen1998new}
Janssen, R.~S., G.~A. Satten, S.~L. Stramer, B.~D. Rawal, T.~R. O'Brien, B.~J.
  Weiblen, F.~M. Hecht, N.~Jack, F.~R. Cleghorn, J.~O. Kahn, et~al. (1998).
\newblock New testing strategy to detect early hiv-1 infection for use in
  incidence estimates and for clinical and prevention purposes.
\newblock {\em Jama\/}~{\em 280\/}(1), 42--48.

\bibitem[\protect\citeauthoryear{Kaplan and Brookmeyer}{Kaplan and
  Brookmeyer}{1999}]{kaplan1999snapshot}
Kaplan, E.~H. and R.~Brookmeyer (1999).
\newblock Snapshot estimators of recent hiv incidence rates.
\newblock {\em Operations Research\/}~{\em 47\/}(1), 29--37.

\bibitem[\protect\citeauthoryear{Kassanjee, McWalter, B{\"a}rnighausen, and
  Welte}{Kassanjee et~al.}{2012}]{kassanjee2012new}
Kassanjee, R., T.~A. McWalter, T.~B{\"a}rnighausen, and A.~Welte (2012).
\newblock A new general biomarker-based incidence estimator.
\newblock {\em Epidemiology (Cambridge, Mass.)\/}~{\em 23\/}(5), 721.

\bibitem[\protect\citeauthoryear{Kassanjee, Pilcher, Busch, Murphy, Facente,
  Keating, Mckinney, Marson, Price, Martin, et~al.}{Kassanjee
  et~al.}{2016}]{kassanjee2016viral}
Kassanjee, R., C.~D. Pilcher, M.~P. Busch, G.~Murphy, S.~N. Facente, S.~M.
  Keating, E.~Mckinney, K.~Marson, M.~A. Price, J.~N. Martin, et~al. (2016).
\newblock Viral load criteria and threshold optimization to improve hiv
  incidence assay characteristics-a cephia analysis.
\newblock {\em AIDS (London, England)\/}~{\em 30\/}(15), 2361.

\bibitem[\protect\citeauthoryear{Konikoff, Brookmeyer, Longosz, Cousins, Celum,
  Buchbinder, Seage~III, Kirk, Moore, Mehta, et~al.}{Konikoff
  et~al.}{2013}]{konikoff2013performance}
Konikoff, J., R.~Brookmeyer, A.~F. Longosz, M.~M. Cousins, C.~Celum, S.~P.
  Buchbinder, G.~R. Seage~III, G.~D. Kirk, R.~D. Moore, S.~H. Mehta, et~al.
  (2013).
\newblock Performance of a limiting-antigen avidity enzyme immunoassay for
  cross-sectional estimation of hiv incidence in the united states.
\newblock {\em PloS one\/}~{\em 8\/}(12), e82772.

\bibitem[\protect\citeauthoryear{Kouyos, von Wyl, Yerly, B{\"o}ni, Rieder,
  Joos, Taff{\'e}, Shah, B{\"u}rgisser, Klimkait, et~al.}{Kouyos
  et~al.}{2011}]{kouyos2011ambiguous}
Kouyos, R.~D., V.~von Wyl, S.~Yerly, J.~B{\"o}ni, P.~Rieder, B.~Joos,
  P.~Taff{\'e}, C.~Shah, P.~B{\"u}rgisser, T.~Klimkait, et~al. (2011).
\newblock Ambiguous nucleotide calls from population-based sequencing of hiv-1
  are a marker for viral diversity and the age of infection.
\newblock {\em Clinical infectious diseases\/}~{\em 52\/}(4), 532--539.

\bibitem[\protect\citeauthoryear{Laeyendecker, Brookmeyer, Cousins, Mullis,
  Konikoff, Donnell, Celum, Buchbinder, Seage~III, Kirk, et~al.}{Laeyendecker
  et~al.}{2013}]{laeyendecker2013HIV}
Laeyendecker, O., R.~Brookmeyer, M.~M. Cousins, C.~E. Mullis, J.~Konikoff,
  D.~Donnell, C.~Celum, S.~P. Buchbinder, G.~R. Seage~III, G.~D. Kirk, et~al.
  (2013).
\newblock Hiv incidence determination in the united states: a multiassay
  approach.
\newblock {\em The Journal of infectious diseases\/}~{\em 207\/}(2), 232--239.

\bibitem[\protect\citeauthoryear{Laeyendecker, Konikoff, Morrison, Brookmeyer,
  Wang, Celum, Morrison, Abdool~Karim, Pettifor, and Eshleman}{Laeyendecker
  et~al.}{2018}]{laeyendecker2018identification}
Laeyendecker, O., J.~Konikoff, D.~E. Morrison, R.~Brookmeyer, J.~Wang,
  C.~Celum, C.~S. Morrison, Q.~Abdool~Karim, A.~E. Pettifor, and S.~H. Eshleman
  (2018).
\newblock Identification and validation of a multi-assay algorithm for
  cross-sectional hiv incidence estimation in populations with subtype c
  infection.
\newblock {\em Journal of the International AIDS Society\/}~{\em 21\/}(2),
  e25082.

\bibitem[\protect\citeauthoryear{Liang and Zeger}{Liang and
  Zeger}{1986}]{liang1986gee}
Liang, K.-Y. and S.~L. Zeger (1986).
\newblock Longitudinal data analysis using generalized linear models.
\newblock {\em Biometrika\/}~{\em 73\/}(1), 13--22.

\bibitem[\protect\citeauthoryear{Mahiane, Fiamma, and Auvert}{Mahiane
  et~al.}{2014}]{mahiane2014mixture}
Mahiane, S.~G., A.~Fiamma, and B.~Auvert (2014).
\newblock Mixture models for calibrating the bed for hiv incidence testing.
\newblock {\em Statistics in medicine\/}~{\em 33\/}(10), 1767--1783.

\bibitem[\protect\citeauthoryear{Maman, Chilima, Masiku, Ayouba, Masson,
  Szumilin, Peeters, Ford, Heinzelmann, Riche, et~al.}{Maman
  et~al.}{2016}]{maman2016closer}
Maman, D., B.~Chilima, C.~Masiku, A.~Ayouba, S.~Masson, E.~Szumilin,
  M.~Peeters, N.~Ford, A.~Heinzelmann, B.~Riche, et~al. (2016).
\newblock Closer to 90--90--90. the cascade of care after 10 years of art
  scale-up in rural malawi: a population study.
\newblock {\em Journal of the International AIDS Society\/}~{\em 19\/}(1),
  20673.

\bibitem[\protect\citeauthoryear{Mastro, Kim, Hallett, Rehle, Welte,
  Laeyendecker, Oluoch, and Garcia-Calleja}{Mastro
  et~al.}{2010}]{mastro2010estimating}
Mastro, T.~D., A.~A. Kim, T.~Hallett, T.~Rehle, A.~Welte, O.~Laeyendecker,
  T.~Oluoch, and J.~M. Garcia-Calleja (2010, Jan).
\newblock {{E}stimating {H}{I}{V} {I}ncidence in {P}opulations {U}sing {T}ests
  for {R}ecent {I}nfection: {I}ssues, {C}hallenges and the {W}ay {F}orward}.
\newblock {\em J HIV AIDS Surveill Epidemiol\/}~{\em 2\/}(1), 1--14.

\bibitem[\protect\citeauthoryear{McDougal, Parekh, Peterson, Branson, Dobbs,
  Ackers, and Gurwith}{McDougal et~al.}{2006}]{mcdougal2006comparison}
McDougal, J.~S., B.~S. Parekh, M.~L. Peterson, B.~M. Branson, T.~Dobbs,
  M.~Ackers, and M.~Gurwith (2006).
\newblock Comparison of hiv type 1 incidence observed during longitudinal
  follow-up with incidence estimated by cross-sectional analysis using the bed
  capture enzyme immunoassay.
\newblock {\em AIDS Research \& Human Retroviruses\/}~{\em 22\/}(10), 945--952.

\bibitem[\protect\citeauthoryear{Moyo, Gaseitsiwe, Mohammed, Pretorius~Holme,
  Wang, Kotokwe, Boleo, Mupfumi, Yankinda, Chakalisa, et~al.}{Moyo
  et~al.}{2018}]{moyo2018cross}
Moyo, S., S.~Gaseitsiwe, T.~Mohammed, M.~Pretorius~Holme, R.~Wang, K.~P.
  Kotokwe, C.~Boleo, L.~Mupfumi, E.~K. Yankinda, U.~Chakalisa, et~al. (2018).
\newblock Cross-sectional estimates revealed high hiv incidence in botswana
  rural communities in the era of successful art scale-up in 2013-2015.
\newblock {\em PloS one\/}~{\em 13\/}(10), e0204840.

\bibitem[\protect\citeauthoryear{Parekh, Kennedy, Dobbs, Pau, Byers, Green, Hu,
  Vanichseni, Young, Choopanya, et~al.}{Parekh
  et~al.}{2002}]{parekh2002quantitative}
Parekh, B.~S., M.~S. Kennedy, T.~Dobbs, C.-P. Pau, R.~Byers, T.~Green, D.~J.
  Hu, S.~Vanichseni, N.~L. Young, K.~Choopanya, et~al. (2002).
\newblock Quantitative detection of increasing hiv type 1 antibodies after
  seroconversion: a simple assay for detecting recent hiv infection and
  estimating incidence.
\newblock {\em AIDS research and human retroviruses\/}~{\em 18\/}(4), 295--307.

\bibitem[\protect\citeauthoryear{Pattanasin, van Griensven, Mock, Sukwicha,
  Winaitham, Satumay, O'Connor, Hickey, Siraprapasiri, Woodring,
  et~al.}{Pattanasin et~al.}{2020}]{pattanasin2020recent}
Pattanasin, S., F.~van Griensven, P.~A. Mock, W.~Sukwicha, S.~Winaitham,
  K.~Satumay, S.~O'Connor, A.~C. Hickey, T.~Siraprapasiri, J.~V. Woodring,
  et~al. (2020).
\newblock Recent declines in hiv infections at silom community clinic bangkok,
  thailand corresponding to hiv prevention scale up: An open cohort assessment
  2005--2018.
\newblock {\em International Journal of Infectious Diseases\/}~{\em 99},
  131--137.

\bibitem[\protect\citeauthoryear{Rehle, Johnson, Hallett, Mahy, Kim, Odido,
  Onoya, Jooste, Shisana, Puren, et~al.}{Rehle
  et~al.}{2015}]{rehle2015comparison}
Rehle, T., L.~Johnson, T.~Hallett, M.~Mahy, A.~Kim, H.~Odido, D.~Onoya,
  S.~Jooste, O.~Shisana, A.~Puren, et~al. (2015).
\newblock A comparison of south african national hiv incidence estimates: a
  critical appraisal of different methods.
\newblock {\em PloS one\/}~{\em 10\/}(7), e0133255.

\bibitem[\protect\citeauthoryear{Sherr, Lopman, Kakowa, Dube, Chawira,
  Nyamukapa, Oberzaucher, Cremin, and Gregson}{Sherr
  et~al.}{2007}]{sherr2007voluntary}
Sherr, L., B.~Lopman, M.~Kakowa, S.~Dube, G.~Chawira, C.~Nyamukapa,
  N.~Oberzaucher, I.~Cremin, and S.~Gregson (2007).
\newblock Voluntary counselling and testing: uptake, impact on sexual
  behaviour, and hiv incidence in a rural zimbabwean cohort.
\newblock {\em Aids\/}~{\em 21\/}(7), 851--860.

\bibitem[\protect\citeauthoryear{Solomon, Mehta, McFall, Srikrishnan,
  Saravanan, Laeyendecker, Balakrishnan, Celentano, Solomon, and Lucas}{Solomon
  et~al.}{2016}]{solomon2016community}
Solomon, S.~S., S.~H. Mehta, A.~M. McFall, A.~K. Srikrishnan, S.~Saravanan,
  O.~Laeyendecker, P.~Balakrishnan, D.~D. Celentano, S.~Solomon, and G.~M.
  Lucas (2016).
\newblock Community viral load, antiretroviral therapy coverage, and hiv
  incidence in india: a cross-sectional, comparative study.
\newblock {\em The lancet HIV\/}~{\em 3\/}(4), e183--e190.

\bibitem[\protect\citeauthoryear{Suligoi, Galli, Massi, Di~Sora, Sciandra,
  Pezzotti, Recchia, Montella, Sinicco, and Rezza}{Suligoi
  et~al.}{2002}]{suligoi2002precision}
Suligoi, B., C.~Galli, M.~Massi, F.~Di~Sora, M.~Sciandra, P.~Pezzotti,
  O.~Recchia, F.~Montella, A.~Sinicco, and G.~Rezza (2002).
\newblock Precision and accuracy of a procedure for detecting recent human
  immunodeficiency virus infections by calculating the antibody avidity index
  by an automated immunoassay-based method.
\newblock {\em Journal of Clinical Microbiology\/}~{\em 40\/}(11), 4015--4020.

\bibitem[\protect\citeauthoryear{Yang, Xia, He, Yang, Ruan, Zhao, Wang, Shao,
  and Pan}{Yang et~al.}{2012}]{yang2012new}
Yang, J., X.~Xia, X.~He, S.~Yang, Y.~Ruan, Q.~Zhao, Z.~Wang, Y.~Shao, and
  X.~Pan (2012).
\newblock A new pattern-based method for identifying recent hiv-1 infections
  from the viral env sequence.
\newblock {\em Science China Life Sciences\/}~{\em 55\/}(4), 328--335.

\end{thebibliography}

\newpage
\setcounter{section}{0}
\renewcommand{\thesection}{S\arabic{section}}
\section{Simulation Procedure}

\subsection{Data Simulation to Mimic \citet{duong2015recalibration}}\label{append:estphi}

In section \ref{sim:external} we describe the process for estimating mean window period and MDRI (and possibly FRR) from an external study.
To get the external study data, we construct a data generating process based off of the data source in \citet{duong2015recalibration}.
The purpose of \citet{duong2015recalibration} was to estimate the MDRI for one particular recency test.
We re-purpose the data source to estimate characteristics of our recency tests 1A-D and 2A-D for the simulation studies.
In \citet{duong2015recalibration}, the authors gathered data from individuals with known seroconversion times (or equivalently infection duration) and with measurements taken longitudinally, sometimes over the course of many years.
In total, we have 2077 longitudinal measurements on 175 individuals.
There were additional individuals in the dataset, but we focused on only the ones with optimal panel data as described in \citet{duong2015recalibration}
The individuals had different HIV-1 subtypes including A, B, C, D, and AE, and were from varied areas of the globe including the Netherlands, Thailand, Ethiopia, Kenya, China, and Trinidad.
We grouped all geographic areas and HIV-1 subtypes together for simplicity.

\begin{figure}[htbp!]
    \centering
    \includegraphics[scale=0.7]{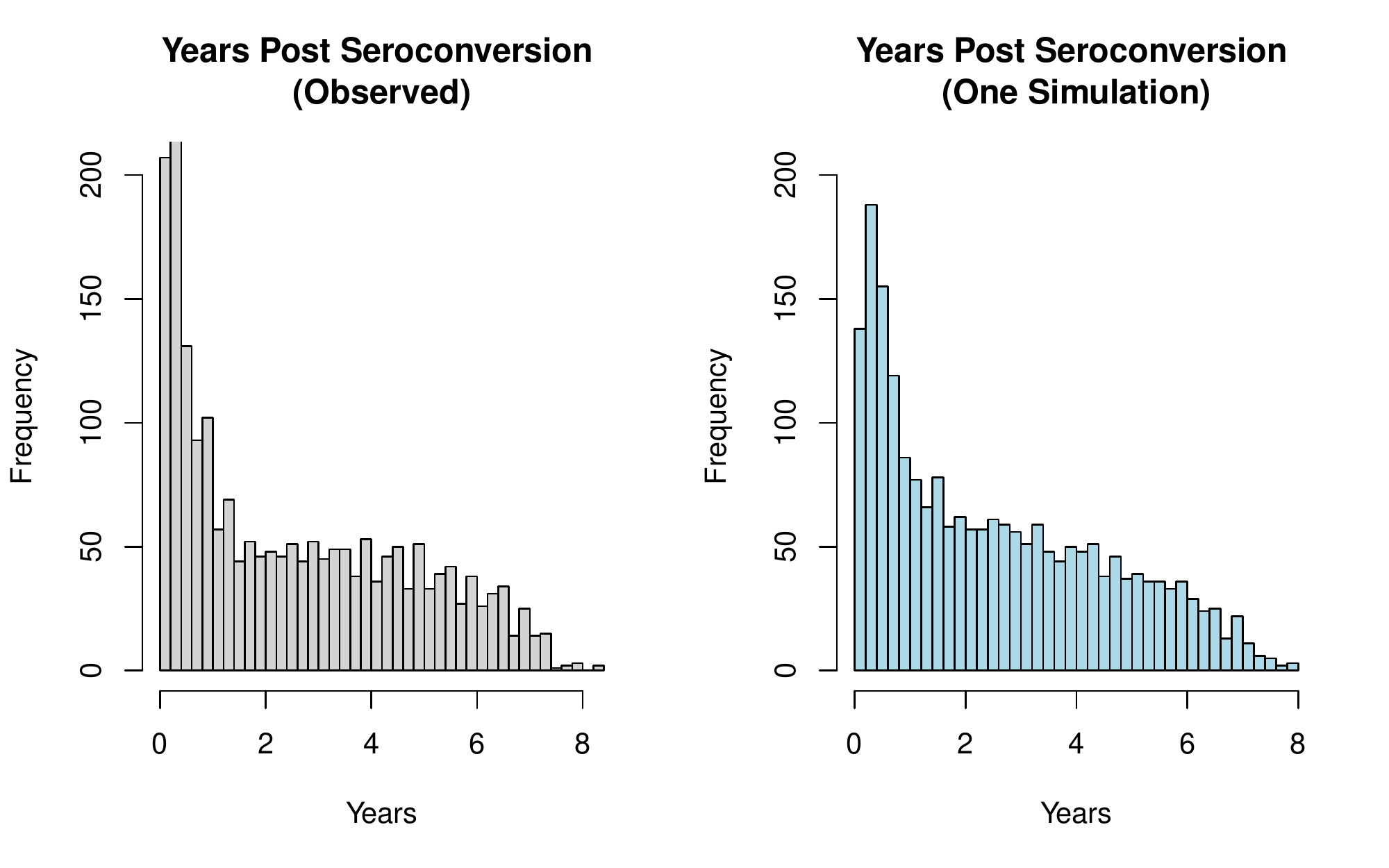}
    \caption{Histograms of year post seroconversion for subject biomarker measurements.
    The observed distribution from \citet{duong2015recalibration} is shown in grey, and one example of a simulated distribution based on the above procedure is shown in blue.}
    \label{fig:duong}
\end{figure}

The grey histogram in Figure \ref{fig:duong} shows the empirical distribution of infection durations in the \citet{duong2015recalibration} dataset pooled over all geographic and subtype cohorts.
Our aim is to create a data generation process that mimics this empirical distribution, but preserves the longitudinal aspect of the data (as opposed to re-sampling independently from this histogram).

\begin{figure}[htbp!]
    \centering
    \includegraphics[scale=0.65]{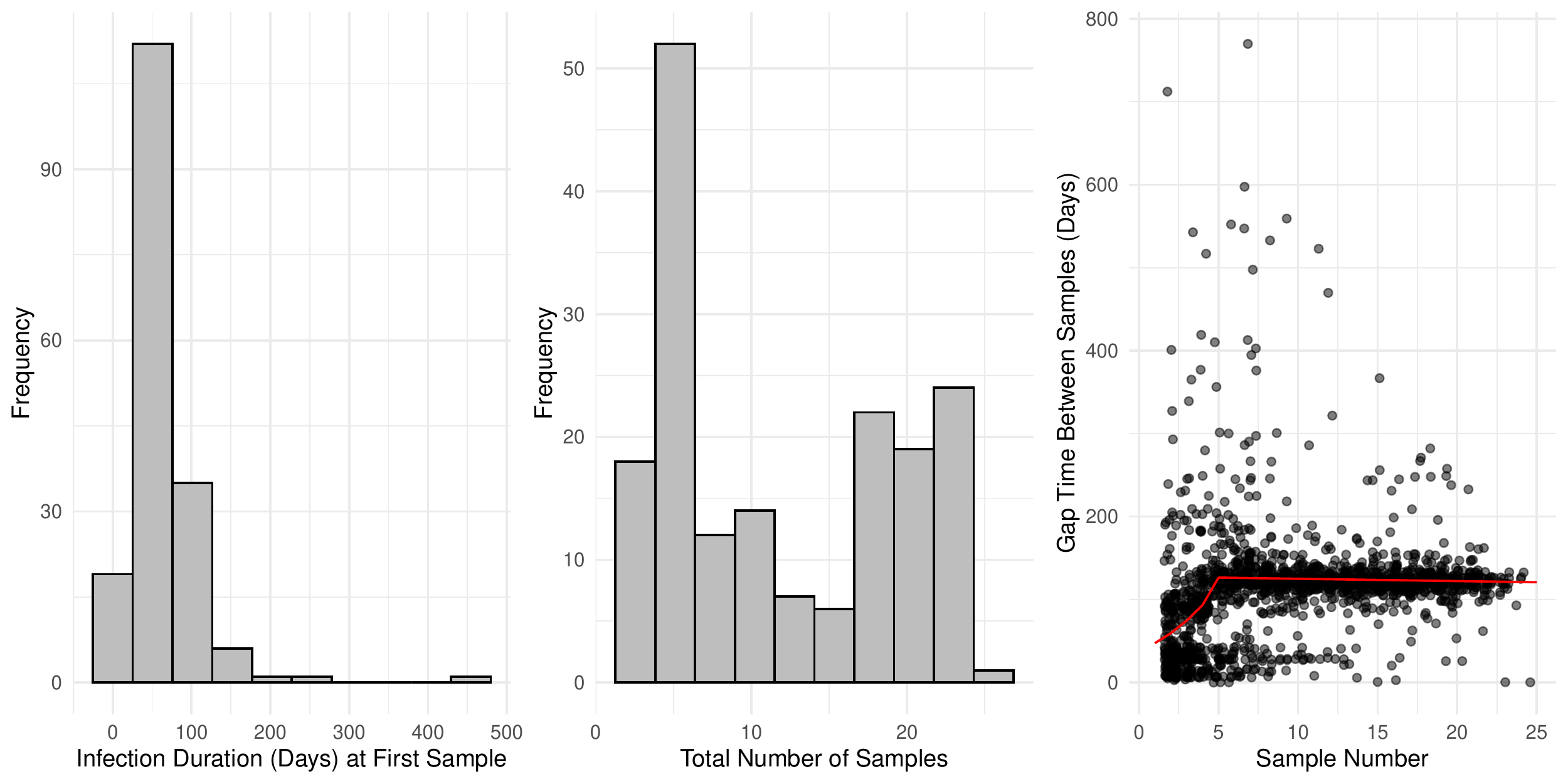}
    \caption{Data from \citet{duong2015recalibration}.
    Left: Histogram of infection duration at first sample.
    Middle: Histogram of total number of longitudinal samples by subject.
    Right: Relationship between sample number and gap days between samples.
    The red line is the fitted mean based on our Poisson GEE regression model (log-linear model with a knot at sample 5).
    The points are jittered to display their relative density.}
    \label{fig:duong-samp}
\end{figure}

For each individual in the original dataset, we calculated the gap times between their longitudinal measurements.
The average number of days between longitudinal measurements is a function of the sample number, i.e., the first couple of samples taken on an individual are typically close together in time and after that there is a longer time between each sample (see Figure \ref{fig:duong-samp}).
Based on this observation, we fit a piece-wise log-linear GEE model with a knot at sample number 5.
The model fit is shown in Figure \ref{fig:duong-samp}.
To simulate one dataset, for each individual we sampled from the empirical distributions of infection duration at first sample and total number of samples (shown in the left and middle panels of Figure \ref{fig:duong-samp}, respectively).
We simulated gap times based on the fitted model for all sample numbers up to the total number.
Finally, we added those gap times to the infection duration at first sample.
The result is shown in the right panel of Figure \ref{fig:duong}.

\subsection{Derivation of Simulation Distributions} \label{append:simudist}

In section \ref{sim:epi}, we described three incidence functions.
In this section, we derive the infection time distributions for each.
To do this, we impose one additional assumption.
\begin{assumpE}\label{ass:simuadd}
$\lambda_t^*(s) = \lambda(s)I\{s\in[t-c_t,t]\}$, where $c_t$ is the solution to
\[\int_{t-c_t}^t\lambda(s)\{1-p(s)\}ds = p(t).\]
In addition, $p_t^*(s) = p(s)$ for $s\in[t-c_t,t]$.
\end{assumpE}
Assumption \ref{ass:simuadd} states that equivalent incidence among eligible populations at different times holds for a range of $s$ closer to $t$, while the incidence $\lambda_t^*(s) = \Pr(T=s|T\ge s,A(t))$ is zero when $s$ is far away from $t$.
It leads to Assumption \ref{ass:approx} if $c\le c_t$ where $c = \tau$ for the snapshot estimator and $c = T^*$ for the adjusted estimator.
This assumption may not generally hold in practice, since we would expect a continuous $\lambda_t^*(s)$, and in this case there may be a discontinuity in incidence at time $t - c_t$, however, it is useful in generating infection time distribution with Assumption \ref{ass:approx} holds.

Based on Assumption \ref{ass:simuadd}, when the prevalence is constant with $p(s)=p$, the infection time $T$ for an HIV-positive subject can be generated as follows.
We first generate $e \sim Unif[0,1]$.
The infection time $T$ is the solution of 
\[\int_{t-c_t}^T\frac{\lambda(s)(1-p)}{p}ds=e.\]
Particularly, the closed form solution for $T$ in a number of cases in shown below.

\begin{table}[htbp]
\begin{centering}
\begin{tabular}{lcc}
\hline\hline
Setting & Incidence $\lambda(s)$ & Infection Time \\\hline
Constant Incidence &$\lambda$ & $t - \dfrac{p(1-e)}{\lambda(1-p)}$\\
Linearly Decreasing Incidence & $\lambda + \rho(t-s)$ ($\rho>0$) & $t - \left(\sqrt{\lambda^2 + 2\rho\dfrac{p}{1-p}(1-e)}- \lambda\right)/\rho$\\
Exponentially Decreasing Incidence & $\lambda \exp\{\rho(t-s)\}$ ($\rho>0$) & $t -  \log\left\{1 + \rho\dfrac{p}{(1-p)\lambda}(1-e)\right\}/\rho$\\
\hline
\end{tabular}
\end{centering}
\end{table}

\begin{figure}[htbp!]
    \centering
    \includegraphics[scale=0.7]{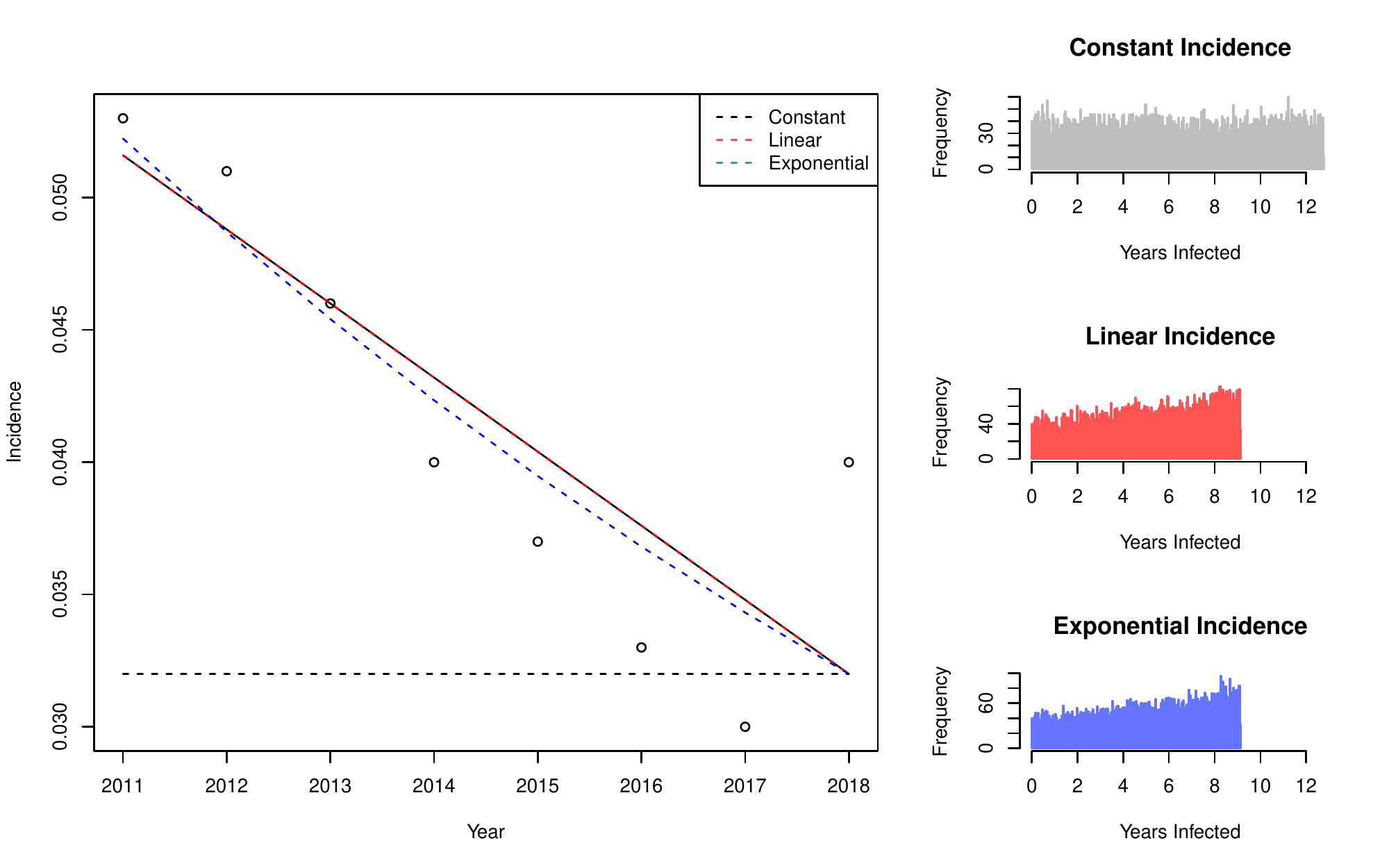}
    \caption{Left: Incidence trends using data from \citet{pattanasin2020recent}: constant incidence, linearly decreasing incidence, and exponentially decreasing incidence.
    We consider all three of these for the simulation settings.
    Right: Distribution of 10,000 past infection times that each setting implies.}
    \label{fig:incidence}
\end{figure}

Figure \ref{fig:incidence} shows the incidence functions we consider for simulations based on the Bangkok MSM data, and the corresponding infection time distributions based on the above derivations.
As a sensitivity check, we slightly varied the parameters derived from the Bangkok data (i.e., prevalence of 0.31 instead of 0.29) and the results did not substantively change.

Note that when we move from constant to linear or exponential incidence, the assumption of uniformly distributed infection times is violated.
Our simulation study assesses how violating this assumption affects the performance of the snapshot and adjusted estimators.

\end{document}